\documentclass[sigconf, authorversion,nonacm]{acmart}

\usepackage{listings}
\usepackage[skins]{tcolorbox}
\usepackage{fancybox}
\usepackage{changepage}
\usepackage{blindtext}
\usepackage{graphicx}
\usepackage{enumitem}
\usepackage{multirow}
\usepackage{booktabs}
\usepackage{subcaption}
\usepackage{url}
\usepackage{soul,xcolor}
\usepackage{color}
\usepackage{siunitx}
\usepackage{makecell}
\usepackage{textcomp}
\usepackage{colortbl}  

\newcommand{\tool}{\textit{MSIVD}\xspace}
\newcommand{\bigvul}{\textit{BigVul}\xspace}
\newcommand{\pb}{\textit{PreciseBugs}\xspace}

\definecolor{deepblue}{rgb}{0,.2,0.6}
\definecolor{deepgreen}{rgb}{0,0.5,0}
\definecolor{deepchampagne}{rgb}{0.98, 0.84, 0.65}
\definecolor{mintgreen}{rgb}{0.6, 1.0, 0.6}
\definecolor{vividviolet}{rgb}{0.62, 0.0, 1.0}
\definecolor{mangotango}{rgb}{1.0, 0.51, 0.26}
\definecolor{dkgreen}{rgb}{0,0.5,0}
\definecolor{dkred}{rgb}{0.5,0,0}
\definecolor{gray}{rgb}{0.5,0.5,0.5}

\lstdefinestyle{javastyle} {
  language=Java,
  basicstyle=\ttfamily\bfseries\footnotesize,
  columns=flexible,
  morekeywords={virtualinvoke},
  keywordstyle=\color{blue},
  ndkeywordstyle=\color{red},
  commentstyle=\color{dkred},
  stringstyle=\color{dkgreen},
  numbers=left,
  breaklines=true,
  numberstyle=\ttfamily\footnotesize\color{gray},
  stepnumber=1,
  numbersep=10pt,
  backgroundcolor=\color{white},
  tabsize=4,
  showspaces=false,
  showstringspaces=false,
  xleftmargin=.23in,
  escapeinside={(*@}{@*)},
}

\colorlet{shadecolor}{gray!40}
\sethlcolor{shadecolor}

\newcommand*\startnumber[1]{%
    \setcounter{lstnumber}{\numexpr#1-1\relax}%
}

\newcommand*\stopnumber{%
    \startnumber{-2}%
}

\begin{document}

\title{Security Vulnerability Detection with Multitask Self-Instructed Fine-Tuning of Large Language Models}

\author{Aidan Z.H. Yang}
\email{aidan@cmu.edu}
\affiliation{
\institution{Carnegie Mellon University}
\city{Pittsburgh}
\country{United States}
}
\author{Haoye Tian}
\email{haoye.tian@unimelb.edu.au}
\affiliation{
\institution{University of Melbourne}
\city{Melbourne}
\country{Australia}
}
\author{He Ye}
\email{hey@cmu.edu}
\affiliation{
\institution{Carnegie Mellon University}
\city{Pittsburgh}
\country{United States}
}
\author{Ruben Martins}
\email{rubenm@cs.cmu.edu}
\affiliation{
\institution{Carnegie Mellon University}
\city{Pittsburgh}
\country{United States}
}
\author{Claire Le Goues}
\email{clegoues@cs.cmu.edu}
\affiliation{
\institution{Carnegie Mellon University}
\city{Pittsburgh}
\country{United States}
}

\begin{abstract}
  Software security vulnerabilities allow attackers to perform malicious activities to disrupt software operations. Recent Transformer-based language models have significantly advanced vulnerability detection, surpassing the capabilities of static analysis based deep learning models. However, language models trained solely on code tokens do not capture either the explanation of vulnerability type or the data flow structure information of code, both of which are crucial for vulnerability detection. 
We propose a novel technique that integrates a multitask sequence-to-sequence LLM with program control flow graphs encoded as a graph neural network to achieve sequence-to-classification vulnerability detection. We introduce \tool, multitask self-instructed fine-tuning for vulnerability detection, inspired by chain-of-thought prompting and LLM self-instruction. Our experiments demonstrate that \tool achieves superior performance, outperforming the highest LLM-based vulnerability detector baseline (LineVul), with a F1 score of 0.92 on the \bigvul dataset, and 0.48 on the \pb dataset. 
By training LLMs and GNNs simultaneously using a combination of code and explanatory metrics of a vulnerable program, \tool represents a promising direction for advancing LLM-based vulnerability detection that generalizes to unseen data. Based on our findings, we further discuss the necessity for new labelled security vulnerability datasets, as recent LLMs have seen or memorized prior datasets' held-out evaluation data.
\end{abstract}


\settopmatter{printacmref=false, printfolios=true}
\maketitle

\section{Introduction}
\label{sec:intro}
Software security vulnerabilities allow attackers to compromise a program and force undesired behaviors, such as exposure of sensitive user information or data extortion. The pervasive threat posed by software vulnerabilities has left a profound impact on individuals and businesses alike~\cite{aslan2023comprehensive}.

This has motivated a long history of prior work to automatically detect such vulnerabilities.  Recent work trains deep learning models for vulnerability detection with information from static analysis, such as on features derived from a program's abstract syntax tree~\cite{devign}, data flow analysis~\cite{deepdfa}, or data dependency analysis~\cite{vuldeepecker, grace}. Including this type of program contextual information improves vulnerability detection accuracy~\cite{deepdfa,ivdetect,grace}. However, deep learning models predicated on static analysis achieve higher precision at a cost to scalability, in terms of the sizes of programs or code blocks that can be considered, and time. For example, IVDetect~\cite{ivdetect} takes up to 9 days to train. The small size of the vulnerability detection datasets available for training also necessarily constrain the performance of these models. 

Recent advances in large language models (LLMs) alleviate both the time and data required for training vulnerability detection models. LineVul~\cite{linevul} achieves state-of-the-art vulnerability detection effectiveness with a fraction of the training time as IVDetect, by leveraging a code pre-trained LLM's prior understanding of code. 
Indeed, the introduction of LLM-based vulnerability detection tools~\cite{codexglue, linevul}, has enabled new techniques that combine combine LLMs with static analysis-based deep learning; these have achieved the highest vulnerability detection rate reported in the prior literature~\cite{deepdfa}. 
However, this effectiveness is still constrained by an LLM's reliance on code-tokens. 
Although recent advances in code pre-trained LLMs have led to a deeper understanding of code semantics~\cite{unnaturalness}, LLMs still struggle to detect vulnerabilities across larger code bases as the number of tokens exceeds their context window size~\cite{llmao}. Cutting larger programs into smaller pieces can mitigate  this challenge~\cite{llmao, linevul}, albeit by discarding data from already relatively small, curated vulnerability datasets. 

Losing information from already fairly small datasets poses a challenge in data-hungry machine learning contexts.  Moreover, we observe that these previously curated vulnerability datasets~\cite{bigvul, precisebugs}  often contain valuable vulnerability information beyond the vulnerable code that is largely unused in state-of-the-art techniques, like an explanation of the vulnerabilities, precise localization information, and a proposed fix.  That is, importantly, these datasets provide insight as to \textit{why} a vulnerability exists at all, and \textit{how} it could be exploited.  Although the datasets are usually fairly small by ML standards, they provide rich information well beyond the code change associated with each vulnerability.  

In this paper, we propose a multitask self-instruction LLM model that trains on multiple dimensions of vulnerability information in combination with dataflow-inspired graph neural networks (GNNs).
Multitask learning enables a model to learn shared knowledge and patterns simultaneously, typically leading to improved generalization and accuracy~\cite{crawshaw2020multi}.
Our proposed tool is based on both recent advances in LLM research that enable fine-tuning on relatively small datasets, and the insights that (1) joint fine-tuning encompassing both code and vulnerability explanations can potentially enhance performance compared to solitary code fine-tuning methods, and (2) most security vulnerabilities entail specific and often subtle information flow, but training language models on either code or explanations alone will not capture  key relations between values and data propagated through a potentially vulnerable program.  Representing the program as a graph is therefore essential.

Inspired by chain-of-thought and self-instruct reasoning~\cite{chainofthought, selfinstruct}, we process labelled vulnerability data into a multi-round dialogue format to fine-tune a self-instruct multitask model. 
We further train our model on program analysis information by adding light-weight graph neural network (GNN) layers with embeddings from control flow graphs (CFG) on top of our fine-tuned LLM.
We first evaluate our model on an established dataset \bigvul. We empirically show that our technique outperforms the previous best-performing LLM-based and static analysis DL-based and 0.17, respectively. 

However, our findings also suggest that modern LLMs exhibit significant evaluation data leakage on established vulnerability datasets. We therefore further pre-process and evaluate on a novel vulnerability dataset using the \textit{PreciseBugsCollector}~\cite{precisebugs}, to ensure that our held-out evaluation dataset only includes code and its labelled vulnerabilities released after our underlying LLM's training cutoff date. We discuss the implications of LLM evaluation data contamination in Section~\ref{sec:results}.

In summary, we make the following contributions.
\begin{itemize}[leftmargin=5mm] 

\item{\textbf{Multitask self-instruct fine-tuning for security vulnerability detection.} We propose a multitask training technique that fine-tunes LLMs on vulnerability detection, vulnerability explanation, and vulnerability fix. We further combine our model with a GNN-based vulnerability adapter to achieve state-of-the-art vulnerability detection effectiveness.}

\item{\textbf{Novel dataset.} Using the \textit{PreciseBugsCollector} \cite{precisebugs}, we collect a labelled vulnerability dataset and pre-process it into a self-instruct dialogue format. To mitigate the potential of LLM data contamination, we filter and evaluate our tool on labelled vulnerabilities from code bases occurring after January 2023, which is the training data cut-off of our pre-trained LLMs. }

\item{\textbf{Empirical evaluation}. We evaluate \tool against state-of-the-art vulnerability detection tools and perform an ablation study.  We show that \tool outperforms baseline across both an established dataset and the novel \pb dataset, and that multi-round self-instruction during fine-tuning is essential to \tool's effectiveness.}

\item{\textbf{Artifact availability}}. Our data, tool, and model checkpoints are available.\footnote{\url{https://zenodo.org/records/11403208}}
\end{itemize}

\section{Illustrative Example}
\label{sec:example}
\lstset{style=javastyle}
\begin{figure}[t!]
\centering
\vspace*{-5mm}
\begin{subfigure}[b]{\columnwidth}
\begin{lstlisting}[numbersep=5pt,xleftmargin=21pt,numberstyle=\scriptsize,basicstyle=\footnotesize\ttfamily,firstnumber=89]
  ...
  if (!checkPassword(userId, password)) {
    throw new RuntimeException( ... ) ; // incorrect password 
  }
(*@ \stopnumber @*) ... (*@ \startnumber{350} @*) 
public boolean checkPassword(String userId, String pword) {
  if (StringUtils.isBlank(userId)) {
    MSException.throwException( ... ); // user ID null
  }
  if (StringUtils.isBlank(pword)) {
    MSException.throwException( ... ); // password null
  }
  UserExample example = new UserExample();
  example.createCriteria().andIdEqualTo(user)
      .andPasswordEqualTo(CodingUtil.md5(pword));
  return userMapper.countByExample(example) > 0;
}
public User selectUser(String userId, String email) {
  User user = userMapper.selectByPrimaryKey(userId);
  if ((user == null) && (StringUtils.isNotBlank(email)) {
    UserExample example = new UserExample();
    example.createCriteria().andEmailEqualTo(email);
    List<User> users = userMapper.selectByExample(example);
    if (!CollectionUtils.isEmpty(users)) {
      return users.get(0);
    }
  }
  return user;
}
    \end{lstlisting}
    \caption{The code snippet with a security vulnerability.}
    \label{code:example_code}
\end{subfigure}
\hfill
\begin{subfigure}[b]{\columnwidth}
\begin{lstlisting}[numbersep=5pt,xleftmargin=21pt,numberstyle=\scriptsize,basicstyle=\footnotesize\ttfamily,firstnumber=1, language=C]
vuln_description = 
/* MeterSphere is an open source continuous testing platform. 
   Version 2.9.1 and prior are vulnerable to denial of service. 
   The `checkPassword` method checks whether the user-provided 
   password matches the password saved in the database. 
   The `CodingUtil.md5` method encrypts the original password
   with MD5 to ensure it is not stored in plain text. 
   If a user submits a very long password, the system is forced
   to execute the long password MD5 encryption process, 
   exausting server CPU and memory, and causing a denial of 
   service attack on the server. */
exploitability_score = 2.8;
severity = "medium";
attack_complexity = "low";
vuln_lines_start = 350;
vuln_lines_end = 375;
\end{lstlisting}
\caption{The vulnerability message.}
\label{code:example_expl}
\end{subfigure}

\vspace*{-2mm}
\caption{\small Example CWE-770 (allocation of resources without limits or throttling) vulnerability. \tool's multi-task fine-tuning uses as features all of the code, vulnerability description, exploitability score, severity, attack complexity, and vulnerable lines.}
\label{code:example}
\Description{example}
\vspace{-0.5cm}
\end{figure}

Figure~\ref{code:example} presents an example to illustrate the insights behind our proposed approach.\footnote{Note that for presentation purposes we have selected an example in Java; we focus our evaluation on defects in C/C++ code, in line with prior work.} The code sample is drawn from a CWE-770 vulnerability\footnote{https://cwe.mitre.org/data/definitions/770.html} from the open source project MeterSphere.\footnote{https://www.cvedetails.com/cve/CVE-2023-32699/}
A CWE-770 vulnerability describes a situation in which a program inadequately limits a resource, in terms of either resource quantity or time dedicated to its use.  
Without adequate use of quotas, resource limits, or other protection mechanisms, an attacker can overwhelm  resources by rapidly making requests, leading to performance degradation or denial of service. 

Figure~\ref{code:example_code} shows the \lstinline{checkPassword} method, which checks if a user has provided in a valid username and a valid password string. On face, the code appears to correctly throw exceptions in response to invalid inputs, providing suitable security checks. However,
the vulnerability description in the CVE (Figure~\ref{code:example_expl}) explains that a malicious user can exhaust server CPU and memory by submitting a very long password, leading to a denial of service.

\begin{figure*}[t!]
\centering
\includegraphics[width=.95\textwidth]{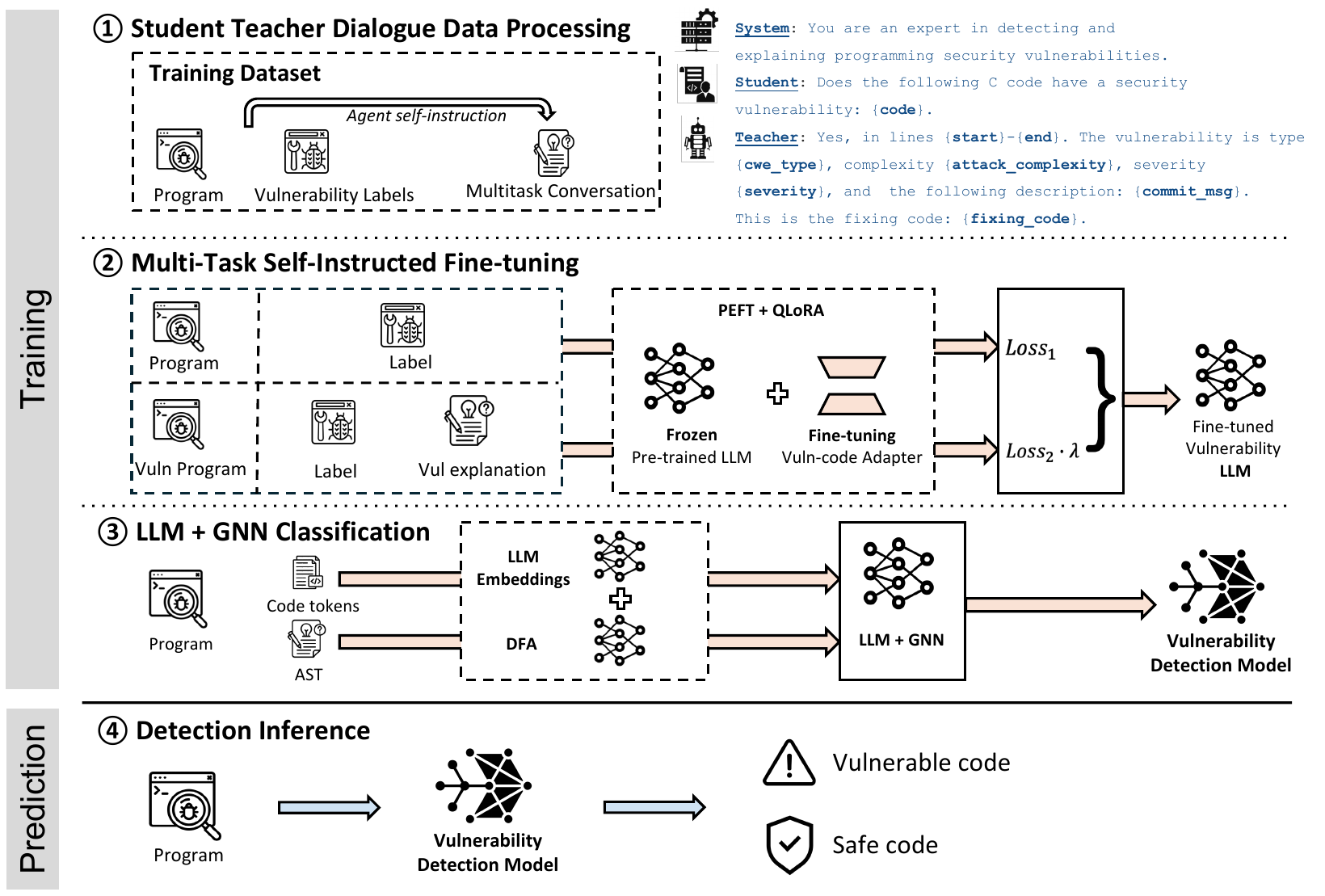}
\caption{\small \tool’s architecture, which takes as training data a code snippet, its vulnerability label, and various human annotated vulnerability labels. \tool outputs a final vulnerability classification on unseen code snippets.}

\label{fig:multivuln_overview}
\Description{overview}
\end{figure*}

This vulnerability is significantly easier to spot given the associated explanation. An untrained human reader benefits from it, as well as the information provided by the CWE type, severity, and complexity information provided by labelled vulnerability datasets. We hypothesize that a large, code-pretrained language-model can also benefit from the additional context provided by explanations like these associated with a known-vulnerable code snippet.

Rich examples like the one in Figure~\ref{code:example} (taken from \pb \cite{precisebugs}) require careful mining of human-annotated vulnerability datasets or other laborious curation, limiting the size of data available for fine-tuning. 
Previously, using relatively small quantities labelled data for instruction-tuned fine-tuning on large language models (i.e., those above 7 billion trainable parameters) was infeasible~\cite{muennighoff2024scaling}. It was more efficient to query the final attention states (which encapsulates an LLM's contextual information for all input elements) of a pre-trained LLM and then performing non-attention based training for vulnerability detection~\cite{llmao, linevul, deepdfa}. Recent advances in LLM fine-tuning have enabled lightweight, parameter efficient~\cite{peft}, and quantized adapter level fine-tuning~\cite{qlora, llmao} suitable for smaller training data~\cite{hsieh2023distilling}. We posit that a combination of the recent advances in LLM fine-tuning, and an argumentation of vulnerability datasets using explanations can improve a language model's understanding of vulnerabilities as a whole.

Finally, notice that \lstinline{checkPassword} (at line 350) is first called by \lstinline{loginLocalMode} at line 90. The code spanning line 90 to line 375 consists of 5179 word tokens, larger than most open source LLM's 2048 or 4096-token context windows. 
If either \lstinline{loginLocalMode} or \lstinline{checkPassword} is used in other contexts beyond line 375, a context window that includes key information about relevant data and information flow grows even larger. This kind of information flow can be derived via dataflow analysis on the program's control flow graph and modeled by a Graph Neural Network (GNN, cf. DeepDFA~\cite{deepdfa}).  
We simultaneously train GNNs with the adaptor weights of a fine-tuned model as our training procedure for vulnerability detection.

\section{Approach}
\label{sec:approach}

\begin{figure}[t!]
\centering
\begin{lstlisting}[numbersep=5pt,xleftmargin=21pt,numberstyle=\scriptsize,basicstyle=\footnotesize\ttfamily,firstnumber=1, language=C, language=C]
Round 0 = {
    role: "System",
    content: "You are an expert in detecting and locating programming security vulnerabilities, and can help answer vulnerability questions",
},
Round 1 = {
    role = "Student",
    content = f"Does the following code have any security vulnerabilities: {code_snippet}",

    role = "Teacher",
    content = f"Yes. The following code has a vulnerability type {cwe_type}.",
},
Round 2 = {
    role = "Student",
    content = f"What is the description of the vulnerablity?",

    role = "Teacher",
    content = f"The vulnerability is: {commit_msg}",
}    
Round 3 = {
    role = "Student",
    content = f"Locate the lines that are vulnerable and should be repaired.",

    role = "Teacher",
    content = f"The code is vulnerable at lines {vuln_lines}, with the following fix: {fixing_code}",
}
\end{lstlisting}
\vspace*{-2mm}
\caption{\small A single training data entry for \tool's vulnerability detection multi-task fine-tuning. The 4 rounds of dialogue between human and bot follows 4 different labelled data: vulnerability classification label, vulnerability description, vulnerability type, and vulnerability repair lines. 
}
\label{code:dialogue}
\Description{dialogue}
\end{figure}

\begin{figure}[t!]
\centering
\includegraphics[width=.45\textwidth]{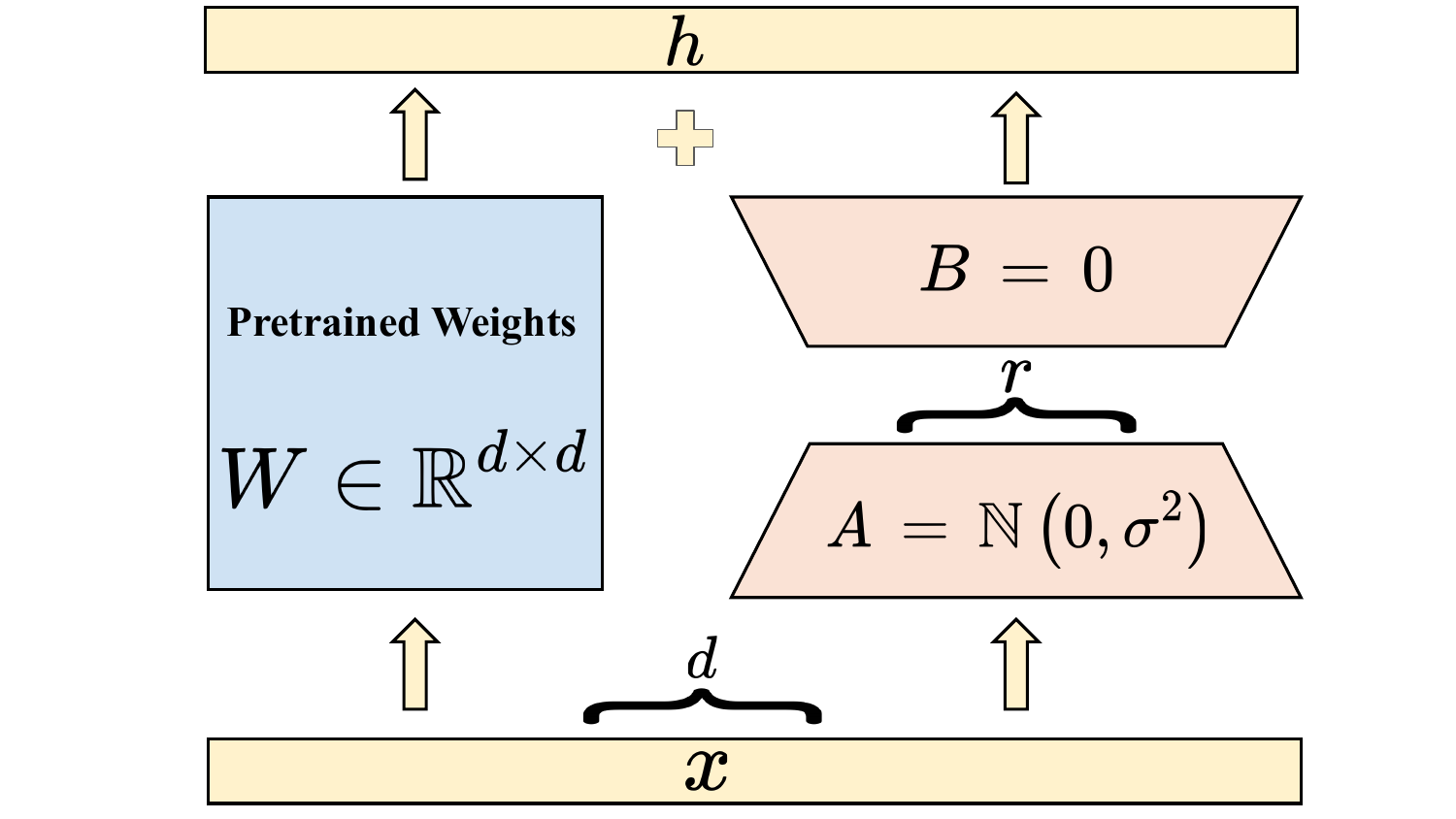}
\caption{\small LoRA re-parameterization for efficient fine-tuning, where only $A$ and $B$ contain trainable parameters, and the initial pre-trained weights $W_0$ remain frozen.}
\label{fig:lora}
\Description{lora}
\end{figure}

Figure \ref{fig:multivuln_overview} provides an overview of \tool, which unfolds in four phases.  The first three phases constitute training.  First, during \textcircled{1} self-instruct dialogue-based data processing, \tool prepares a given dataset for fine-tuning by extracting vulnerability characteristics including type, description, and source code location (Section~\ref{sec:phase1})
The second step, \textcircled{2} multi-task fine-tuning,  uses multi-task learning to fine-tune a LLM, targeting two learning objectives: 
(1) detecting a vulnerability and (2) providing an explanation that describes the vulnerability's characteristics.  Section~\ref{sec:phase2} provides more detail.
The third step, \textcircled{3} LLM+GNN training, jointly trains the LLM and a GNN based on information flow data derived from the program's control flow graph (Section~\ref{sec:phase3}). 
In the \textcircled{4} detection phase, given a program, the vulnerability detection LLM trained by \tool predicts whether a program contains a vulnerability (Section~\ref{sec:phase4}).

\subsection{Student Teacher Dialogue Data Processing}
\label{sec:phase1}

Model training includes a code snippet, its associated vulnerability label, CWE-type, a vulnerability description (e.g., how an attacker could exploit the vulnerability), and developer fix with fix location. Inspired by chain-of-thought reasoning~\cite{chainofthought}, we process the vulnerable code and labels into a multi-round conversation format between a teacher and student. Inserting intermediate reasoning steps improves the ability of a LLM to perform complex reasoning~\cite{chainofthought}. Embedded with the conversation is first a system prompt asserting that the teacher is ``an expert in detecting and explaining programming security vulnerabilities'', followed by a back-and-forth of questions and answers. The teacher-learner chain-of-thought learning is based on Self-instruct~\cite{selfinstruct} and Dialogue-policy-planned~\cite{deng2023plug}.
Each complete dialogue is a single training data entry.

Figure \ref{code:dialogue} shows a complete dialogue training data entry example. The teacher and student converse in three rounds of dialogue, each on a different aspect of the security vulnerability in a target code snippet. The first round of dialogue discusses the existence of the vulnerability; the second round, an explanation of why the code snippet has a vulnerability; and the third, which lines needed to be changed to fix the vulnerability. Figure~\ref{code:example} shows examples for the \lstinline{code_snippet}, \lstinline{cve_type}, \lstinline{complexity} and \lstinline{commit_msg} variables inserted into the dialogues, discussed in Section~\ref{sec:example}.

To produce non-vulnerable samples from our dataset, we sample developer-fixed code from the dataset, and associate it with a negative label. Specifically, we create a single-round dialogue where the \lstinline{Teacher} tells the \lstinline{Student} that ``the code does not have a security vulnerability''.

\subsection{Multi-Task Self-Instructed Fine-Tuning}
\label{sec:phase2}

We follow the approach proposed by MFTCoder~\cite{mftcoder} to make full use of the produced self-instruct dialogue dataset by targeting multiple objectives simultaneously via multi-task training.
Specifically, we use a multi-turn conversation approach, launching
two agents, one acting as ``teacher'', and the other as ``student''.
As shown in Figure \ref{code:dialogue}, the first round of dialogue concerns the existence of a vulnerability in the code sample. 
Under the hood of the training process, the LLM repeatedly generates teacher dialogue outputs to answer the student's questions, and compares against the ground-truth answer to calculate loss.

For fine-tuning efficiency, we use Parameter-efficient fine-tuning (PEFT)~\cite{peft} and Quantized Large-scale Language Model Low-Rank Adaptation (QLoRA)~\cite{qlora} with 4-bit quantization.
QLoRA incorporates a high-precision quantization technique (NF4), quantizes the pretrained model to 4-bits, and trains a small set of lightweight adapter weights, based on Low-Rank Adaptation (LoRA)~\cite{lora}. 

Figure~\ref{fig:lora} describes the key idea behind LoRA, which is 
to insert trainable rank decomposition matrices into each layer of the Transformer architecture, reducing the number of trainable parameters.
Figure~\ref{fig:lora} shows a forward pass from dense layer $x$ to dense layer $h$, where $d$ is the initial rank, and $r$ is the lower-ranked LoRA  adapter. 

During fine-tuning, the pre-trained neural network weight matrix $W \in \mathbb{R}^{d \times d}$ remains fixed, and only the dimensional expansion matrix $A \in \mathbb{R}^{r \times x}$ and $B \in \mathbb{R}^{d \times r}$ undergo training. If $W_0$ are the initial model weights, LoRA modifies the forward pass to:
\begin{equation}W_0 + \Delta W = W_0 + BA\end{equation}

Figure~\ref{fig:lora} shows a random Gaussian initialization for $A$, and zero for $B$, so $\Delta W = BA$ is zero at the beginning of training. 

To ensure the convergence of loss across multiple training tasks, we use the weighted loss calculation proposed by MFCoder~\cite{mftcoder}:

\begin{equation}
\mathcal{L}(\theta) = \min_\theta \frac{1}{N} \sum_{i=1}^{N} \frac{\sum_{j=1}^{M_i} \sum_{k=1}^{T_{ij}} -\log(p_0(t_{ijk}))}{\sum_{j=1}^{M_i} T_{ij}}
\end{equation}

Where \(N\) represents the total number of tasks, \(M_i\) denotes the number of samples for the \(i\)-th task, \(T_{ij}\) is the count of valid tokens involved in the loss function for the \(j\)-th sample of the \(i\)-th task, and \(t_{ijk}\) is the \(k\)-th valid token of the \(j\)-th sample for the \(i\)-th task.
Equation 2 effectively takes the average loss across all $N$ tasks.

\subsection{LLM + GNN Classification}
\label{sec:phase3}

Our multi-task fine-tuning outputs a sequence-to-sequence model trained strictly on code and word tokens. Program information flow is also often important to detect security vulnerabilities. We therefore additionally represent programs as graph embeddings to learn information propagation patterns. Graph learning starts with an initial representation of a node, and then performs a fixed number of iterations of the message-passing algorithm to learn a graph's message-passing patterns~\cite{gilmer2017neural}. 
This constitutes an additional modelling phase that ultimately results in  
sequence-to-classification vulnerability detection (i.e., a binary output indicating vulnerability presence/absence). 

We use dataflow analysis (DFA) embeddings to set up a GNN model pipeline, as inspired by DeepDFA~\cite{deepdfa}.  DeepDFA's abstract dataflow embedding aims to directly represent variable definitions propagated in a program's control flow graph (CFG), allowing a GNN to learn a dataflow analysis algorithm. 
Specifically, DeepDFA performs a \textit{reaching definition analysis} over a program's CFG to compute, for each program point, which variable definitions can reach that program point. A variable definition reaches a program point if there exists a path in the CFG between that definition and that point, without a redefinition of that same variable along that path.
DeepDFA's DFA embeddings 
use the Gated Graph Sequence Neural Network (GGNN)~\cite{li2015gated},
where the GNN \textit{aggregation} of information from all nodes is a Multi-Layer Perceptron (MLP) and information \textit{update} on each node is a Gated Recurrent Unit (GRU).

To apply this approach in our context, we use the GNN as a light-weight adapter layer, and concatenate its learned embeddings at each training iteration with the hidden states of our fine-tuned LLM along its last dimension. The last hidden states of a LLM encapsulate the information for all input elements before model prediction. By concatenating the embeddings during each forward pass, we can train a LLM and a GNN simultaneously. 
To ensure that the combined model leverages the prior instruction-tuned weights, we convert ``0'' and ``1'' integer-labels into tokenized ``yes'' and ``no'' string-labels, and apply a \textsc{LogSoftmax} layer to obtain a final list of logits, which we convert into binary classification.
Unlike DeepDFA, our approach is the first to concatenate embeddings from an instruction-tuned LLM with a GNN to perform classification, and customize a loss function that enables training stability.

\subsection{Detection inference}
\label{sec:phase4}

For a given program, \tool queries the trained vulnerability detection model to predict a code snippet as vulnerable or safe. Since the model was previously instruction-tuned on vulnerability explanations in a conversational format, it can also be prompted to provide a code snippet's specific CWE type, and explanation of why a vulnerability exists.  We focus in this paper on evaluating vulnerability detection accuracy, and leave an evaluation of explanation quality for future work.

\section{Evaluation Setup}
\label{sec:setup}

In this section, we describe our evaluation setup, including our datasets (Section~\ref{sec:dataset}),  metrics and baselines (Section~\ref{sec:baselines}), and our model setup (Section~\ref{sec:model}).

\subsection{Datasets}
\label{sec:dataset}

\begin{table*}[h]
    \centering
    \begin{tabular}{llrr}
        \toprule
        \textbf{Vulnerability Type} & \textbf{CWE Examples} & \textbf{\bigvul (\%)} & \textbf{\pb (\%)} \\
        \midrule
        Buffer Error           & CWE-125, CWE-787 & \num{33626} (19.8\%)  & \num{3547} (27.3\%)  \\
        Input Validation Error &CWE-134, CWE-89   & \num{22867} (13.5\%)  & \num{1761} (13.6\%)  \\
        Resource Error         &CWE-415, CWE-404  & \num{30270} (17.8\%)  & \num{2756} (21.2\%)  \\
        Privilege Escalation   &CWE-264, CWE-255  & \num{29485} (17.4\%)  & \num{145} (8.8\%)   \\
        Value Error            & CWE-190, CWE-369 & \num{13628} (8.0\%)   & \num{1367} (10.5\%)  \\
        Other                  &CWE-434, CWE-122  & \num{39896} (23.5\%)   & \num{2394} (18.5\%)  \\
        
        \midrule
        Total Samples              && \num{169772} & \num{12970}  \\
        Mean \# lines per sample    && \num{30} & \num{377}  \\
        Total Lines of Code    && \num{4530522} & \num{4615582}  \\\midrule 
        Total Vulnerabilities  && \num{3754}    & \num{2543} \\
        \bottomrule
    \end{tabular}
    \caption{Datasets used in our evaluation.  \bigvul is the focus of evaluation in prior work.  We also use the \emph{PreciseBugCollector}~\cite{precisebugs} to collect a novel dataset of C/C++ vulnerabilities reported after the cut-off training date for the considered LLMs.}
    \label{table:cwe_types}
\end{table*}

\subsubsection{Established Vulnerability  Dataset}
Recently-proposed vulnerability detection models~\cite{deepdfa, ivdetect, linevd, linevul} are typically evaluated on the \textit{Devign}~\cite{devign} or \bigvul~\cite{bigvul} datasets. Both contain real-world C/C++ projects and vulnerabilities. We choose \bigvul for our evaluation because \bigvul is equipped with code snippets (in the form of single functions), labels, and CWE explanations, while \textit{Devign} only provides code snippets and labels. 
Furthermore, \bigvul is larger than \textit{Devign} (\textit{Devign} has 14,653 labelled functions, and \bigvul has 169,772). Unlike \textit{Devign}, \bigvul is an imbalanced dataset, consisting of 94\% non-vulnerable labels and 6\% vulnerable labels. 
Following prior work, we split \bigvul into a 80/10/10 split on training, evaluating, and testing.

Following LineVul~\cite{linevul} and DeepDFA~\cite{deepdfa}, we excluded a total of 1,564 labelled functions (0.8\%) from the Big-Vul dataset, namely samples with
(1) incomplete functions (i.e., ending with ‘);’, or not ending in
‘\}’) that cannot be parsed for dataflow analysis
(2) functions where no lines were added or removed, but were simply labelled vulnerable. (3) functions where more than 70\% of lines are modified for the fix, indicating a substantial change that may fundamentally change the vulnerable code, and 
(4) functions that are fewer than 5 lines long.

\subsubsection{Novel Vulnerability Dataset}
\bigvul contains vulnerabilities sampled from \textit{before} most modern LLM's training cut-off date of January 2023~\cite{bigvul, devign, linevul}. Since our tool is based on pre-trained LLMs, we aim to also collect labelled vulnerability data produced after that date, to mitigate the risk of data leakage. 
We use the \textit{PreciseBugCollector}~\cite{precisebugs} toolset to produce this dataset.
\textit{PreciseBugCollector} mines verified vulnerabilities reported by human annotators from the National Vulnerability Dataset (NVD), which includes significant and well-known vulnerabilities, such as HeartBleed (CVE-2014-0160\footnote{\url{https://nvd.nist.gov/vuln/detail/cve-2014-0160}}), Spectre (CVE-2017-5753 and CVE-2017-5715\footnote{\url{https://nvd.nist.gov/vuln/detail/cve-2017-5753} and \url{https://nvd.nist.gov/vuln/detail/cve-2017-5715}}), and Log4Shell (CVE-2021-44228\footnote{\url{https://nvd.nist.gov/vuln/detail/cve-2021-44228}}).
\textit{PreciseBugCollector} uses the NVD API\footnote{\url{https://nvd.nist.gov/developers/vulnerabilities}} to download comprehensive vulnerability metadata. This produces a rich overall dataset of human expert-confirmed vulnerabilities 
accompanied by information like vulnerability descriptions, types identified by the CWE (Common Weakness Enumeration) ID, severity levels, and references, often including source and patches. 

We filter the 217,403 vulnerabilities constituting the entire NVD database 
by identifying those with external links that lead to GitHub commits and are tagged with \textit{Patch}. This produces 9,759 vulnerabilities with code.  We extract the fixed source code from the corresponding GitHub commits. We split the 9,759 vulnerabilities into individual file patches, producing 47,302 file changes.\footnote{Our tool, like prior work, assumes that vulnerabilities are localized to single files, which is not always the case; we discuss limitations to this assumption in Section~\ref{sec:threats}.}

Although \pb includes vulnerabilities in C, C++, Python, Java, Go, Rust, and PHP, our initial multitask training results indicated that \pb across multiple languages induces too much noise for stable training. This is in line with prior results: multi-language fine-tuning on security vulnerabilities significantly diminishes performance from single-language fine-tuning~\cite{instruct_vul}. We therefore further filter \pb for only C/C++ code, in line with \bigvul. This produces a dataset of 2543 vulnerabilities, which we split into 12,970 code-snippet samples by single file changes.  

Our \pb dataset consists of 80\% non-vulnerable and 20\% vulnerable samples. We purposefully craft our dataset to be unbalanced to make it more difficult for a model to guess (i.e., 50\% coin toss) the correct answer, and to replicate real world settings (most code in the real world is not vulnerable). 
We split the dataset into a 80/10/10 split on training, evaluating, and testing.
Instead of randomly splitting, we create our evaluation and testing datasets by filtering only for vulnerabilities where the associated vulnerability label and code fix occurred after January 2023.
For our training dataset, we keep vulnerability samples from all \textit{PreciseBugCollector} before January 2023.

Our contribution to the original \pb dataset is in (1) splitting the samples into single file code snippets to fit into LLM context windows, (2) processing each sample into a student-teacher dialogue format, and, most importantly, (3) re-running \textit{PreciseBugCollector} on the most recent 2023-2024 labelled vulnerabilities to mitigate LLM evaluation data leakage.

\subsubsection{Analysis}

Table~\ref{table:cwe_types} characterizes the datasets. Following \bigvul and prior work, we primarily report analysis by sample, where a vulnerability is typically comprised of multiple samples (annotated functions or context window-sized code segments). \bigvul contains a much larger sample size (169,772) compared to our collected C and C++ \pb dataset (12,970). However, we note that each sample of \bigvul contains a single program function, with a mean of \textbf{30} lines. Each sample of \pb consists of a program window, which may or may not be contained within a single function, with a mean of \textbf{356} lines. By total lines of code and total number of vulnerabilities, \pb  is comparable to \bigvul in size.

We label each sample's main vulnerability type based on their labelled CWE-type, following Steenhoek et al.'s classification scheme~\cite{steenhoek2023empirical}.
The main vulnerability types are buffer overflow, input validation
error, resource error, privilege escalation, and value error.
Table~\ref{table:cwe_types} shows that the \pb and \bigvul distributions are similar.  

\subsection{Metrics and Baselines}
\label{sec:baselines}

For classification, we convert the existence of a vulnerability into binary labels. To characterize classification effectiveness for an entire
dataset, we use F1, precision, and recall: 
$F1 = \frac{TP} {TP + 0.5(TP + FN)}$, $Precision = \frac{TP}{(TP + FP)}$, 
$Recall = \frac{TP}{TP + FN}$.
In all definitions, TN is
true negative, TP is true positive, FP is false positive, and FN is false negative.
We use the same metrics as prior vulnerability detectors~\cite{deepdfa, ivdetect, linevul, vuldeepecker} for fair comparison.

We compare \tool to baselines across several categories:

\begin{itemize}[leftmargin=5mm]
\item\emph{\textbf{Non-LLM} deep learning-based vulnerability detection tools}: We compare evaluation effectiveness directly to
VulDeePecker~\cite{vuldeepecker}, SySeVR~\cite{codet5}, Draper~\cite{draper}, IVDetect~\cite{ivdetect}, and DeepDFA~\cite{deepdfa}.
\item\emph{\textbf{LLM}-based approaches}: We evaluate the vulnerability detection of open-source, pre-trained LLM models CodeBERT~\cite{codebert}, CodeT5~\cite{codet5}, and CodeLlama~\cite{codellama}. 
We also compare to LineVul~\cite{linevul}, which trains an additional sequence-classification model on top of a pre-trained LLM. LineVul originally uses CodeBERT and RoBERTA~\cite{roberta} as its pre-trained LLM. For a fair comparison, we customize LineVul to use the same pre-trained model, CodeLlama-13B, as \tool. Otherwise, any differences between \tool and LineVul could be a due to a difference in pre-trained model effectiveness, instead of the actual approach.
\item\emph{\textbf{LLM + GNN} combined techniques}: We use DeepDFA's replication package and customize DeepDFA to combine their GNN embeddings with our fine-tuned model, and any HuggingFace\footnote{https://huggingface.co/models} pre-trained model directly. We release customized version of LineVul and DeepDFA that works with all HuggingFace pre-trained models for future research.\footnote{https://zenodo.org/records/11403208}
\item\emph{\textbf{Random}:} We include a random baseline that predicts whether a sample is vulnerable with a probability of 0.5.  We include this baseline to ground the precision, recall, and F1 scores, where performance is sensitive to the underlying data distribution (and our datasets are imbalanced).  
\end{itemize}

The chosen baseline tools represent the state-of-the-art of vulnerability detection
models~\cite{steenhoek2023empirical}, and all work directly on the \bigvul dataset.
Only the LLM models work directly with newer datasets, as they do not require extant program analysis results on dataset code; prior non-LLM tools were designed for \bigvul. We can therefore not evaluate them on the \pb dataset. To evaluate prior LLM tools on vulnerabilities released after their training data cutoff date, we evaluate the top-performing \bigvul LLMs on our collected \pb dataset. 

\subsection{Model setup}
\label{sec:model}

We train two models for \tool. The first performs a sequence-to-sequence fine-tuning of a selected pre-trained LLM, using our multitask self-instruct approach. The second performs a sequence-to-classification training loop that outputs a binary classification label (i.e., if the sample is vulnerable or not), which we build on top of DeepDFA's GNN architecture.
The second model takes the final hidden states from the frozen in place first model. We refer to the tool using both models as \tool throughout evaluation; the tool consisting only of the first model, without the GNN architecture, as \tool$^{-}$. \tool$^{-}$ converts the first model into a sequence-to-classification model directly, using a single \lstinline{linear} layer.
For the initial pre-trained model, we use CodeLlama-13B-Instruct~\cite{codellama}, which is the 13 billion parameters instruction tuned version of CodeLlama. CodeLlama released 4 model size versions, from 7B to 70B. Due to limited computing and VRAM, we chose the 13B version. 

Table~\ref{table:hyperparams} shows the hyperparameters used for both models. The 4352 model dimension from the LLM-GNN model is a result of concatenating the fine-tuned LLM (4096) with the GNN model (256). Similarly, we add the output layers of LLM with the GNN to form $8 + 3 = 11$ layers.
For batch size, we use 4 to fit CodeLlama 13B onto a single RTX 8000 GPU. However, other GPUs with more VRAM could employ higher batch sizes for greater efficiency.

\begin{table}[t]
\caption{Hyperparameters used for multitask self-instruct fine-tuning, and the LLM-GNN combined vulnerability detection model training.
}
\begin{tabular}{l|rr}
\toprule
\textbf{Hyperparameter} & Multitask FT & LLM+GNN \\
\midrule
Initial Learning Rate  &  1e-5 &  1e-6\\
Model Dimension & 4096 & 4352\\
Context Window & 2048 & 2048 \\
Layers  &  8 & 11\\
Batch Size & 4 & 4 \\
Epochs & 10 & 5 \\
\bottomrule
\end{tabular}
\label{table:hyperparams}
\end{table}

\section{Results}
\label{sec:results}

In this section, we present the results evaluating \tool's performance by answering our three research questions:

\begin{enumerate}[leftmargin=3em]
    \itemsep0em 
    \item[\textbf{RQ1:}] How effective is \tool for finding vulnerabilities on established datasets?     We evaluate \tool on \bigvul and compare its vulnerability detection effectiveness to prior baselines. (Section~\ref{sec:rq1})
    \item[\textbf{RQ2:}] To what extent can \tool generalize to unseen vulnerabilities?     Since \tool is based on an underlying LLM that may have already seen the \bigvul dataset, we further evaluate \tool on a novel dataset, \pb. (Section~\ref{sec:rq2})
    \item[\textbf{RQ3:}] How does each component of \tool impact its performance? We aim to discover each component's impact on vulnerability detection. We also evaluate \tool performance on more specific vulnerability types. (Section~\ref{sec:rq3})
\end{enumerate}

All results presented in this section were obtained using an Intel(R) Xeon(R) 6248R CPU @ 3.00GHz running Debian GNU/Linux 1 and two Nvidia Quadro RTX 8000 GPUs.

\subsection{RQ1: How effective is \tool for finding vulnerabilities on established datasets?}
\label{sec:rq1}

\begin{table}[t!]
  \centering
\caption{Vulnerability prediction effectiveness on the BigVul and PreciseBugs datasets. 
VulDeePecker, SySeVR, Draper, and IVDetect performance on BigVul taken from the IVDetect paper~\cite{ivdetect}. CodeBERT and CodeT5 performances on BigVul taken from the DeepDFA replication package~\cite{deepdfa}.
For fair comparison, we customized both LineVul and DeepDFA's replication package to use CodeLlama, and train for the same number of epochs as \tool (5 epochs). 
We also include a random approach that predicts a sample as vulnerable with a probability of 0.5. \tool$^{-}$ is \tool without the GNN.
}
\begin{tabular}{p{1cm}|p{1cm}|p{1.6cm}rrr}
\toprule
 \textbf{Dataset} & \textbf{Type} & \textbf{Technique} &
 \textbf{F1} & \textbf{Precision} & \textbf{Recall}\\
\midrule
\multirow{9}{*}{BigVul}

&\multirow{1}{*}{Random}
&  Random & 0.11  & 0.06  & 0.50   \\
\cmidrule{2-6}

&\multirow{4}{*}{\parbox{0.9cm}{Non-LLM}}
& VulDeePecker & 0.12  & 0.49  & 0.19   \\
&& SySeVR & 0.15  & 0.74  & 0.27   \\
&& Draper& 0.16  & 0.48 & 0.24   \\
&& IVDetect& 0.23  & 0.72 & 0.35   \\
&& DeepDFA & 0.67  & 0.54 & 0.90   \\
\cmidrule{2-6}

&\multirow{3}{*}{LLM}
& CodeBERT              & 0.21  & 0.68  & 0.13   \\
&& CodeT5               & 0.46  & 0.56  & 0.39   \\
&& CodeLlama            & 0.74  & 0.85  & 0.63 \\
&& LineVul  & 0.81  & 0.86  & 0.78   \\
\cmidrule{2-6}
&\multirow{2}{*}{\parbox{0.9cm}{LLM + GNN}}
& CodeT5 + DeepDFA    & 0.79  & 0.85  & 0.71   \\
&& LineVul + DeepDFA  & 0.88  & 0.88  & 0.89   \\
&& \cellcolor{black!25} \tool          &  \textbf{0.92} &  \textbf{0.93} &\textbf{ 0.91 }  \\
\midrule

\multirow{1}{*}{\parbox{1.5cm}{Precise Bugs}}
&\multirow{1}{*}{Random}
&  Random & 0.29  & 0.20  & 0.50   \\
\cmidrule{2-6}

&\multirow{1}{*}{LLM}
&   CodeLlama   & 0.22  & 0.16  & 0.35   \\
&& LineVul & 0.31  & \textbf{0.43} & 0.25 \\
&& \cellcolor{black!25} \tool$^{-}$   & \textbf{0.48}  & 0.40  &\textbf{0.57}   \\

\bottomrule
\end{tabular}
\label{table:multift_results}
\end{table}

\paragraph{Results.} Table~\ref{table:multift_results} shows the effectiveness of our tool on the \bigvul dataset, as well as prior baselines. We re-ran LineVul and DeepDFA, and used results on VulDeePecker, SySeVR, Draper, and IVDetect from the IVDetect paper~\cite{ivdetect}.
DeepDFA's data flow analysis-based GNN technique outperforms prior Non-LLM techniques, with a F1 score of 0.67. DeepDFA's greatest improvement over prior non-LM techniques is via its high recall score of 0.9 (it correctly identifies 90\% of the vulnerable code samples).

Table~\ref{table:multift_results} also shows that all LLM approaches other than CodeBERT perform better than program-analysis DL based approaches. LineVul customized with CodeLlama achieves a F1 of 0.81. That is, LineVul, without insights from program analysis, surpasses all state-of-the-art program analysis based deep learning tools. 

That said, LineVul using CodeLlama, combined with DeepDFA, yields even higher   vulnerability detection effectiveness. We see that when an underlying LLM (i.e., CodeLlama) can already achieve a high F1 score, further model knowledge from static analysis provides limited improvements. In comparison, a more dated LLM like CodeT5 benefits from static analysis more (i.e., a F1 score improvement from 0.46 to 0.79). LLM-based detectors' results (e.g., 0.81 F1 from LineVul) suggest that recent LLMs have a high degree of confidence on vulnerability detection on code tokens alone.

Overall, \tool yields a F1 score of 0.92, precision 0.93, and recall 0.91, outperforming all other baselines on all metrics. The improved results from \tool show that the different aspects of vulnerability explanation can provide further detection accuracy to a pre-trained LLM. However, \tool only shows incremental improvements over LineVul + DeepDFA, as compared to LineVul + DeepDFA's larger improvements on all non-LLM tools. The largest improvements on vulnerability detection with the \bigvul dataset comes from the underlying LLM itself. 

\begin{tcolorbox}
[colback=white,colframe=black,arc=0pt,boxrule=0.5pt,title=RQ1 Summary,boxsep=2pt,left=1pt,right=1pt,top=1pt,bottom=1pt,fonttitle=\bfseries] 
LLM-based techniques outperform non-LLM techniques on the \bigvul dataset. \tool outperforms prior state-of-the-art LLM-based model LineVul with a F1 score of 0.92. The incremental improvements of adding either GNNs or fine-tuning suggests that the underlying pre-trained LLM is capable at vulnerability prediction based on code tokens alone.
\end{tcolorbox}

\subsection{RQ2: To what extent can \tool generalize to unseen vulnerabilities?}
\label{sec:rq2}

\paragraph{Setup.} To measure \tool's ability to generalize to unseen vulnerabilities, we evaluate \tool on the \pb dataset with a January 2023 cut-off date, and compare against LineVul (the best-performing prior technique) as baseline. Note that we use DeepDFA's GNN model as the basis of the implementation of \tool's GNN layer, and DeepDFA's GNN model was set up for \bigvul specifically. Therefore, for the \pb dataset, we use \tool$^{-}$, without the GNN adapter layers. We evaluate the contribution of individual components of \tool's design in Section~\ref{sec:rq3}.

\paragraph{Results.} Table~\ref{table:multift_results} shows \tool's performance on the \pb dataset. \tool shows a larger improvement from LineVul on the newer dataset (compare \tool's F1 score of 0.48 to LineVul's F1 score of 0.31), showing greater effectiveness of our fine-tuning approach on unseen vulnerabilities.

To gather insight as to why LineVul with CodeLlama as the underlying model performs so much better on the \bigvul dataset than the \pb dataset, we measure CodeLlama's effectiveness on our evaluation data directly (i.e., directly using CodeLlama's output logits for prediction). As seen in Table~\ref{table:multift_results}, CodeLlama achieves an F1 score of 0.74 on \bigvul, but only 0.22 on \pb. Without any additional vulnerability classification training or sequence-to-sequence fine-tuning, CodeLlama already beats most prior non-LLM techniques (see, for example, DeepDFA's F1 score of 0.67). 

\begin{figure}
\vspace*{-2mm}
\centering
\includegraphics[width=0.95\columnwidth]{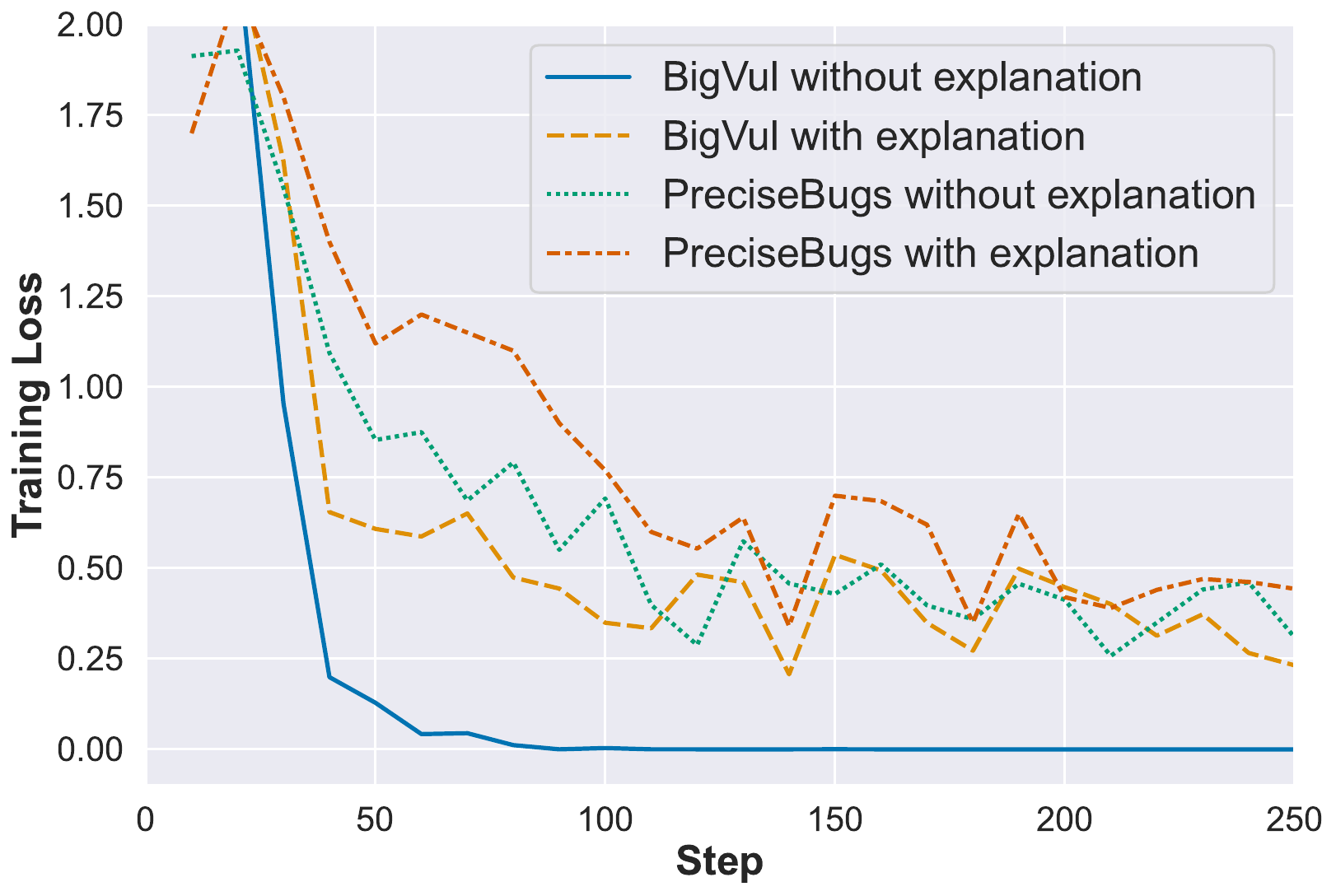}
\caption{Loss curve on \tool with \bigvul and \pb. A lower loss value indicates model predictions that are closer to the ground-truth labels, and a near-zero loss indicates over-fitting.
Note that we also run the exact experiment on the \textit{Devign} dataset, and observe the same loss curve as \bigvul without explanation. }
\label{fig:loss}
\Description{loss}
\end{figure}

These trends are supported by an inspection of the training loss function for CodeLlama using our multitask fine-tuning method.
Figure~\ref{fig:loss} shows the loss curves of our training approach on the \bigvul dataset with (i.e., multitask  fine-tuning) and without explanations (i.e., label-only fine-tuning). A deep-learning model's loss curve describes how closely a model's predictions are to the ground truth. Lower loss means better prediction. The loss curve on fine-tuning \bigvul with explanation approaches 0.2 in 400 steps (2 epochs, roughly 16 hours of training time). In contrast, the loss curve on fine-tuning \bigvul without explanations approaches 0.2 in 50 steps (1/4 of an epoch, roughly 2 hours of training time). Based on our findings, we posit that training an LLM-based model on labelled vulnerabilities released before modern LLM's training cut-off date exhibits {\bf over-fitting} (i.e., LLM prior memorization of the dataset). However, instruction-tuning on vulnerability explanations is much less overfit.

These results support the importance of evaluating LLM-based models on newer labelled vulnerabilities released after the selected LLM's training data cut-off date. We also recommend fine-tuning an LLM on previously seen data with a multitask approach, inserting a higher degree of randomness in its learning.

\begin{tcolorbox}
[colback=white,colframe=black,arc=0pt,boxrule=0.5pt,title=RQ2 Summary,boxsep=2pt,left=1pt,right=1pt,top=1pt,bottom=1pt,fonttitle=\bfseries]
Neither CodeLlama nor prior LLM-based vulnerability detector baselines generalize well to the unseen \pb dataset. Training loss curves suggest that CodeLlama has likely memorized the \bigvul dataset rather than learning to generalize from it. While \tool is more effective on \bigvul compared to \pb vulnerabilities, we confirm that \tool better generalizes to the recently released \pb dataset than prior baselines. 
\end{tcolorbox}

\subsection{RQ3: How does each component of \tool impact its performance?}
\label{sec:rq3}

\begin{table}[t!]
  \centering
\caption{\tool ablation study. ``Pre-trained" uses the underlying LLM directly for vulnerability detection. ``Label-only fine-tuned'' (FT) performs single-task fine-tuning on the vulnerability classification labels. ``Single round self-instruct fine-tuned (SIFT)'' trains the LLM without the agent explanation multi-round dialogue. ``Multi-round SIFT'' uses multi-task agent-dialogue fine-tuning (\tool$^{-}$). ``Multi-round SIFT + GNN'' adds the GNN adapter layer and corresponds to the full version of \tool.
}

\begin{tabular}{p{1.5cm}|lrrr}
\toprule
 \textbf{Dataset} & \textbf{Technique} &
 \textbf{F1} & \textbf{Precision} & \textbf{Recall}\\
\midrule
\multirow{3}{*}{\bigvul}
& Pre-trained & 0.74  & 0.85  & 0.55   \\
& Label-only FT & 0.71 & 0.77  & 0.66    \\
& Single-round SIFT & 0.81  & 0.86  & 0.61   \\
& Multi-round SIFT      & 0.90  & 0.91  & 0.87 \\
& \parbox{2cm}{Multi-round SIFT + GNN} & 0.92  & 0.93 & 0.91   \\
\midrule

\multirow{3}{*}{\pb}
& Pre-trained & 0.22  & 0.16  & 0.35   \\
& Label-only FT & 0.31  & 0.43  & 0.25   \\
& Single-round SIFT & 0.33  & 0.46  & 0.25   \\
& Multi-round SIFT  & 0.48  & 0.4   & 0.57\\
\midrule
\multirow{3}{*}{\parbox{2cm}{\pb Vuln. Type}}
& \tool$^{-}$ Input  & 0.46  & 0.49   & 0.44    \\
& \tool$^{-}$ Resource  & 0.58  & \textbf{0.63}   & 0.51    \\
& \tool$^{-}$ Buffer  &\textbf{ 0.59}  & 0.62   & \textbf{0.57 }   \\

\bottomrule
\end{tabular}
\label{table:multift_ablation}
\end{table}

\paragraph{Setup.} To answer RQ3, we evaluate \tool under four settings and evaluate their performances on \bigvul and \pb. First, we use the underlying pre-trained LLM directly for prediction. We then use a fine-tuned version of \tool, but without any vulnerability explanations in its training data (label-only FT). Finally, we include vulnerability explanations in a single round of self-instruction fine-tuning (single-round SIFT) and multiple rounds of self-instruction fine-tuning (multi-round SIFT, which corresponds to \tool$^-$). For \bigvul, we also add the GNN adapter layers (multi-round SIFT + GNN, which corresponds to the full version of \tool). 

We additionally evaluate our tool on specific vulnerability types within \pb, (training/evaluating on single vulnerability types). We choose the three most common types from our \pb dataset: buffer error (27.3\% of \pb), resource error (21.2\% of \pb), and input validation error (13.6\%). 

\paragraph{Results.} Table~\ref{table:multift_ablation} shows results on both the \bigvul and \pb datasets. As discussed in Section~\ref{sec:rq2}, CodeLlama already performs well at detecting \bigvul vulnerabilities. Training a separate model without using agent self-instruction slightly improves effectiveness, with a F1 score of 0.81 (+0.07 above the pre-trained F1 score of 0.74) for \bigvul, and a F1 score of 0.33 (+0.1 above pre-trained) on \pb.

Surprisingly, we find that fine-tuning on only the vulnerability label and none of the explanations actually performs worse than using a pre-trained model directly for vulnerability classification (0.71 F1 for fine-tuned CodeLlama, and 0.74 F1 for pre-trained CodeLlama on the \bigvul dataset). 
Our findings are consistent with those of Yusuf et al.~\cite{instruct_vul}, who observed that instruction-based fine-tuning may not always enhance performance, especially across a dataset of diverse CWEs. The shift from sequence-to-sequence fine-tuning to the sequence-classification training within a small dataset may simply include more noise, reducing classification performance.

Fine-tuning with both code and vulnerability explanations with the multitask agent setup (\tool) yields the highest vulnerability detection on both \bigvul and \pb. 
We also see that training with multi-round SIFT yields higher F1 scores than single-round SIFT (a F1 improvement of 0.09 for \bigvul, and 0.02 for \pb), which is consistent with prior work on LLM instruction-prompting~\cite{zero-shot}. Finally, we observe that the additional GNN (multi-round SIFT + GNN) provides an additional 0.02 F1 on top of multi-round SIFT for the \bigvul dataset. The incremental improvement from the addition of GNN shows that CodeLlama already makes accurate predictions based on prior knowledge on the \bigvul dataset, as previously discussed in Section~\ref{sec:rq2}.

Table~\ref{table:multift_ablation} shows that when we train and evaluate on single vulnerability types, the F1 scores are improved across all three vulnerability types as compared to results on the entirety of \pb. However, the much smaller training dataset comes with the trade-off of higher precision but lower recall. 

These results further corroborate that the LLM-unseen vulnerabilities in the newer \pb dataset are more difficult for any language model to detect. However, our results do indicate that training with a multi-round self-instruct format on a dataset with both label and explanation produces considerable improvements over pre-trained models alone.

\begin{tcolorbox}
[colback=white,colframe=black,arc=0pt,boxrule=0.5pt,title=RQ3 Summary,boxsep=2pt,left=1pt,right=1pt,top=1pt,bottom=1pt,fonttitle=\bfseries]
Further training a code LLM on vulnerability-specific code and labels improves detection effectiveness. Fine-tuning an LLM without vulnerability explanations actually reduces effectiveness as compared to the pre-trained model. Multitask fine-tuning with all included vulnerability explanations achieves the highest detection effectiveness, especially with multiple rounds of self-instruction. Finally, selecting specific vulnerability types for both training and evaluating yields higher F1 scores, but with a trade-off of lower recall due to the smaller data size.
\end{tcolorbox}

\section{Threats}
\label{sec:threats}
Our evaluation against prior state-of-the-art vulnerability detector tools relies on the \bigvul dataset, as this dataset is supported by all our chosen baseline tools. \bigvul is imbalanced and can more accurately represent a diverse set of real-world vulnerabilities as compared to the \textit{Devign} dataset~\cite{devign, bigvul, deepdfa}. However, \bigvul’s data collection is solely based on bug-fixing commits, which can lead to label noise and selection bias. \bigvul's collected bug-fixing commits are also from GitHub repositories from before most modern LLM's training data cutoff date, leading to evaluation data contamination. To mitigate these issues, we use \textit{PreciseBugCollector}~\cite{precisebugs}, which uses a combination of bug-fixing commits and bug injections to create a similar dataset. 
Our custom data collector filters the evaluation and testing dataset to only use code changes from January 2023 and onwards, which is the training data cutoff for all our selected LLMs. By evaluating our tool and baseline on the larger and time-filtered vulnerability specific \pb dataset, we can reduce the risk of label data noise and evaluation data contamination.

Our choice of vulnerability representation is a program slice for the \pb dataset, and a single function for the BigVul dataset. In practice, real world vulnerabilities typically span across multiple program slices or methods. Sejfia et al.~\cite{sejfia2024toward} coined the term \textit{multiple base units} (MBU) vulnerabilities to describe vulnerabilities that span across different code locations within a repository. Sejfia et al.~\cite{sejfia2024toward} found a 37\% of the function-level vulnerabilities in \bigvul were actually MBU vulnerabilities. To address this issue, we collect our \pb dataset with randomized windows around either bug injections or code-change commits, which could span across multiple functions or classes. However, a more accurate representation of real world vulnerabilities detection would include a prediction for entire MBU vulnerabilities, which we leave as future work.  

\section{Related Work}
\label{sec:related}

\paragraph{Vulnerability Detection}
Security vulnerability detection has a rich research history covering both static and dynamic techniques (see., e.g., ref~\cite{sokMemory,secVulSurvey,fuzzingSurvey}); we focus on work using machine learning for static vulnerability detection, as it is most closely related to ours and recent results are especially promising.
Devign~\cite{devign}, IVDetect~\cite{ivdetect}, and LineVD~\cite{linevd} used GNN on a program's AST or CFG to learn a program's likelihood to be vulnerable. LineVul~\cite{linevd} queried the final attention layers of a language model (CodeBERT) for the specific purpose of vulnerability detection. 
VulDeePecker~\cite{vuldeepecker} trained a BiLSTM on data dependencies for vulnerability detection.
LLMAO~\cite{llmao} is the first LLM-based tool to focus on line-level (as opposed to file level or method level) vulnerability localization.
DeepDFA~\cite{deepdfa} uses DFA algorithms to train a GNN and achieve state-of-the-art vulnerability detection results more efficiently than prior program analysis based tools. Steenhoek et al.~\cite{deepdfa} observed that DeepDFA, when combined with LineVul~\cite{linevd}, yields a higher vulnerability detection effectiveness than all prior Transformer-based tools. Our work shows that this approach can further improve through fine-tuning the underlying LLM on vulnerability-specific data.
Our results also highlight the important risk of data leakage in evaluating LLM-based vulnerability detection techniques.  

The most closely related work to ours is Yusef et al.~\cite{instruct_vul}, which is the first to study the impact of natural language instructions on improving vulnerability detection. However, Yusef et al.~\cite{instruct_vul} only used classification labels for training and did not use multi-task self-instruct training.
Our work is the first to augment training data with vulnerability labels as vulnerability explanations, and train a fine-tuned code-based LLM with a GNN adapter that encodes program DFA results.

\paragraph{LLM-based Agents for code}
An LLM-based code agent uses an LLM for its operation or decision-making process on coding tasks (e.g., automated program repair).
SWE-Agent~\cite{yang2024sweagent} built input and output formats to enable an LLM to view, edit and execute code files. SWE-Agent~\cite{yang2024sweagent} introduced the Agent-Computer Interface (ACI) for repairing Github issues. VERT~\cite{vert} leveraged testing and verification techniques to create an LLM-based transpilation agent.
CodeAgent~\cite{zhang2024codeagent} is a LLM-based agent framework that directly calls external tools for repository level code generation. CodeAgent~\cite{zhang2024codeagent} evaluated their tool by integrating five programming tools for information retrieval, code navigation, and code testing. Our work does not use an LLM directly as an agent for decision-making, but instead processes training data into an agent-learner conversation format. This agent-learner conversation format is able to incorporate labelled software vulnerability data into a multitask fine-tuning model. 

\paragraph{Multitask Fine-Tuning}
Significant prior work builds on the idea of multitask learning for various tasks~\cite{crawshaw2020multi, mftcoder, pascal2021maximum}. T5~\cite{raffel2020exploring} explores applications of multitask learning and transfer learning on LMs; ExT5~\cite{aribandi2021ext5} improved upon T5 by increasing the number of tasks, with mutual interference between tasks. NLG-LM~\cite{zhu2019multi} is a multitask training method for natural language generation. In addition to generating correct and complete responses, NLG-LM explicitly targets response naturalness.

The most similar work to ours is MFTCoder~\cite{mftcoder}, which uses Self-instruct~\cite{selfinstruct}, a custom multitask loss function, and PEFT QLoRA~\cite{peft, qlora} for efficient fine-tuning on competitive coding assignment problems.
We adopt a similar approach but specifically set up for vulnerability explanations. 
Our work is the first multitask tool to use guide a model with vulnerability explanations, and include embeddings from a program analysis-inspired GNN for transfer learning.

\section{Conclusion}
\label{sec:conclusion}
Automatically detecting software security vulnerabilities is a rich and longstanding problem.  
Recent advances in ML have enabled techniques that either combine program analysis with deep learning, or apply LLMs directly to detect vulnerabilities.  
Meanwhile, the relatively small curated security vulnerability datasets provide rich additional information that prior work has left unexplored.  Bridinging this gap, in this paper, we introduce a self-instruct based multitask fine-tuning model to learn vulnerability classification based on both program code and vulnerability explanations. We further include information from data flow analysis, and build a light-weight GNN adapter based on a program's call graph to achieve simultaneous transfer-learning between LLM and GNN. 
Our tool surpasses prior state-of-the-art results on established vulnerability datasets. Furthermore, because of the risk (and evidence) of LLM data contamination, we collect a novel vulnerability dataset with evaluation and test samples  exclusively filtered to be vulnerabilities identified past our pre-trained code LLM's training cutoff, and show that our technique outperforms prior work on that dataset as well.  
Note that we have built our tool on only open-source LLMs to support future reproducibility and extensibility, and make our artifact available with data, tool, and model checkpoints.\footnote{\url{https://zenodo.org/records/11403208}}

\bibliographystyle{ieeetr}
\bibliography{ref}

\end{document}